\documentclass[lettersize,journal]{IEEEtran}
\usepackage{amsmath,amsfonts}
\usepackage[caption=false,font=normalsize,labelfont=sf,textfont=sf]{subfig}
\usepackage{stfloats}
\usepackage{url}
\usepackage{verbatim}
\usepackage{graphicx}
\usepackage{lineno,hyperref}
\usepackage{algorithmic}
\usepackage{textcomp}
\usepackage{booktabs}% http://ctan.org/pkg/booktabs
\usepackage{array}
\usepackage{makecell}
\usepackage{multirow}
\usepackage{tabularx}
\usepackage{times}
\usepackage{color,soul}
\usepackage{fancyhdr,graphicx,amsmath,amssymb}
\usepackage[ruled,vlined]{algorithm2e}
\begin{document}

\title{Unmanned Aerial Vehicles Traffic Management Solution Using Crowd-sensing and Blockchain}

\author{Ruba Alkadi, Abdulhadi Shoufan
\\
C2PS, Khalifa University, UAE

 }       % <-this % stops a space
%\thanks{C2PS, Khalifa University,UAE}}% <-this % stops a space
%\thanks{Manuscript received April 19, 2021; revised August 16, 2021.}

%\author{Abdulhadi Shoufan,~\IEEEmembership{C2PS, Khalifa University,UAE}}

\markboth{IEEE Transactions on Network and Service Management}%
{Shell \MakeLowercase{\textit{et al.}}: A Sample Article Using IEEEtran.cls for IEEE Journals}


\maketitle

\begin{abstract}
Unmanned aerial vehicles (UAVs) are gaining immense attention due to their potential to revolutionize various businesses and industries. However, the adoption of UAV-assisted applications will strongly rely on the provision of reliable systems that allow managing UAV operations at high levels of safety and security. Recently, the concept of UAV traffic management (UTM) has been introduced to support safe, efficient, and fair access to low-altitude airspace for commercial UAVs. A UTM system identifies multiple cooperating parties with different roles and levels of authority to provide real-time services to airspace users. However, current UTM systems are centralized and lack a clear definition of protocols that govern a secure interaction between authorities, service providers, and end-users. The lack of such protocols renders the UTM system unscalable and prone to various cyber attacks. Another limitation of the currently proposed UTM architecture is the absence of an efficient mechanism to enforce airspace rules and regulations. To address this issue, we propose a decentralized UTM protocol that controls access to airspace while ensuring high levels of integrity, availability, and confidentiality of airspace operations. To achieve this, we exploit key features of the blockchain and smart contract technologies. In addition, we employ a mobile crowdsensing (MCS) mechanism to seamlessly enforce airspace rules and regulations that govern the UAV operations. The solution is implemented on top of the Etheruem platform and verified using four different smart contract verification tools. We also provided a security and cost analysis of our solution. For reproducibility, we made our implementation publicly available on Github\footnote{https://gist.github.com/rubaalkadi/820d8aeb015aa4f67f0c496bf051be8d}.
\end{abstract}


\section{Introduction}

The emergence of Unmanned Aerial Vehicles (UAVs), also known as drones, has enabled a wide range of applications in the smart city context. World-class companies like Amazon and Google have realized this potential and started investing in this technology \cite{Amazon,googleWing}. The aviation industry has as well shown a great interest in UAV-based applications. Goods delivery, urban air mobility (UAM), and UAV-as-a-service (UAVaaS) are examples of applications that drive the research in this field. Yet, such applications are far from commercial adoption. That is, several safety and security challenges are limiting the public acceptance of the integration of drones in the urban airspace. It is undeniable that the expected volume of traffic in the urban airspace will present a real threat to public privacy, safety, and security. Particularly, common threats such as spying, physical collisions, and carrying explosives are limiting the public acceptance of the drone \cite{AYDIN2019101180}. Traditional systems for Air Traffic Management (ATM) are not prepared to meet the required level of autonomy and mobility exhibited by the unmanned air traffic \cite{rumba}. Governments worldwide are striving to enable a safe and secure air traffic ecosystem that boosts the public acceptance of low-altitude urban air traffic while ensuring a sufficient level of autonomy and mobility of UAVs. 

Towards this goal, civil aviation agencies in many countries have initiated multiple UAV Traffic Management (UTM) projects \cite{mohamed2018preliminary, ChinaUOM, barrado2020u}. The National Aeronautics Space Agency (NASA) has probably provided the most comprehensive and up-to-date version of the UTM architecture \cite{NASAUTMConOps}. For consistency, we adopt the terminology introduced in the NASA-UTM proposal for the rest of this article. 

The ultimate goal of a UTM system is to orchestrate the roles of stakeholders and individuals involved in the deployment of UAV-based applications including international and national legislation organizations, UAV service providers (USS),  UAV operators, end-users, insurance companies, law enforcement, and the public. Yet, it is articulated that the UTM regulation and infrastructure are still "lagging behind" the latest technological innovations \cite{rumba}. In particular, the current UTM proposal does not support fully autonomous and beyond-line-of-sight (BVLOS) flights which are at the core of most of the proposed applications such as UAM and goods delivery. Rather, a central authority is responsible to handle UAV registration and authorization using traditional time-consuming procedures. Another issue of the current UTM system is the absence of a monitoring system that enforces the introduced rules and regulations in the urban vicinity. Although the literature is rich in contributions that describe technologies for UAV detection \cite{azari2018key, taha2019machine}, tracking, and interdiction \cite{guvenc2018detection}, most of these solutions are expensive and non-scalable. Alternatively, Remote Identification (RID) has been introduced into the UTM system to allow the public to identify the drone and its operator. Accordingly, UAVs need to continuously broadcast a special message that contains relevant information such as a unique identifier, the UAV location, and a timestamp \cite{abdalla2020machine}. This identification mechanism is used to ensure that the public is able to transparently identify the mission of the flights above them which in turn is expected to improve the public acceptance of the UTM. Nonetheless, the current RID scheme does not address concerns regarding possible cybersecurity breaches such as the authenticity of the RID and the confidentiality of the mission information. Moreover, this identification approach suffers from from several issues related to useability, security, and scalability. For example, the affected public can hardly assess the legitimacy of a drone flight which makes people hesitate to report. Also, drones can usually fly at high speeds and people may feel less motivated to report a drone that has just bypassed and disappeared. On the other hand, this human-centered approach is unscalable for law enforcement and can hardly help to identify false or malicious reports.

Despite the great efforts made to modify the current UTM architecture, the security aspect is barely touched in the recent concept of operation \cite{NASAUTMConOps}. Particularly, the proposal does not outline a clear protocol to monitor and penalize unlawful activities. Instead, the USA's Federal Aviation Administration (FAA) holds the full responsibility for enforcing rules and regulations and monitoring operators' compliance as stated in the recent version of the UTM Concept of Operations \cite{NASAUTMConOps}. This centralization introduces more pressure on the central aviation authority and makes the system prone to a single point of failure. Finally, the current Concept of Operations requires the USS to archive all the logs related to all flights which is technically inefficient and non-scalable.

\begin{figure}
    \centering
    \includegraphics[width = \linewidth]{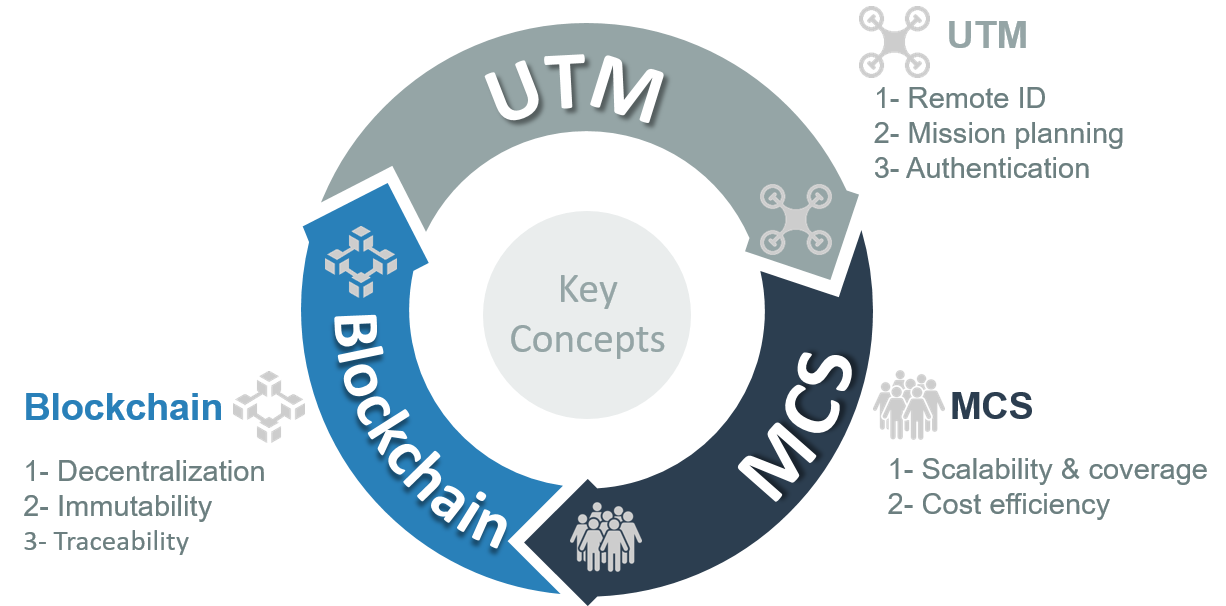}
    \caption{Key concepts in the proposed system. }
    \label{keyfeatures}
\end{figure}

In this work, we address these limitations of the current UTM system by deploying state-of-the-art technologies, namely blockchain and crowdsensing. Specifically, we introduce a fully distributed UTM system that tackles the security issues in the current system. We, further, integrate a crowdsensing scheme to monitor operating UAVs and enforce the UTM regulations. We orchestrate the two technologies within the UTM context to enable an efficient, secure, and scalable UTM system that supports fully autonomous and BVLOS operations. The key features of our proposed solution are illustrated in Figure \ref{keyfeatures}. First, we employ the main concepts introduced in the UTM architecture including Remote ID, mission planning, and authentication which is provided by an authorized party such as the aviation authority. These concepts are deployed on a blockchain-based framework to exploit its key features such as decentralization, immutability, and traceability. The MCS technology is employed within the blockchain framework to ensure that all airspace operators abide with the rules, regulations, and mission plans provided by the UTM system. 

Our work is thus distinguished by the following aspects: 
\begin{itemize}
    \item 	We propose and implement a novel smart contract-based protocol that securely governs the UTM operations. Our solution supports autonomous mission scheduling and deconflicting while preserving high levels of security, safety, and privacy. 
    \item We propose and implement an MCS-based solution to enforce airspace rules and regulations. Further, we introduce an incentive mechanism to enhance the performance of the MCS scheme. 
    \item We analyze and consolidate our solution by a detailed security and cost analysis. We also verify our implemented model using two verification tools. 
\end{itemize}

The rest of the paper is organized as follows: Section \ref{background} provides a background on the components of our solution. Section \ref{literature} surveys the literature and highlights our contributions. Section \ref{overview} describes the proposed model on a high level, and section \ref{implementation} provides the implementation details and algorithms. Sections \ref{test}, \ref{Performance}, and \ref{security} tests, validates, and analyzes the performance and security of the implemented solution, respectively. Finally, Section \ref{conclusion} concludes the paper. 

\section {Background}\label{background}

In this section, we briefly introduce the basic concepts related to our proposed system to help the reader comprehend our contribution. We start by describing the UTM system and the associated remote ID framework. We, then, briefly introduce the blockchain technology as well as the emerging MCS mechanism.  

\subsection{NASA UTM}

To enable safe, efficient, and fair access to low-altitude airspace for UAVs, NASA has been repeatedly updating its draft of the UTM architecture \cite{kopardekar2016unmanned}. In this architecture, the authors identified three main agents: UAV operator/owner, UAV service suppliers (USS), and the regulator, which they referred to as the air navigation service provider (ANSP). 
\begin{figure}[ht]
    \centering
    \includegraphics[width=\linewidth]{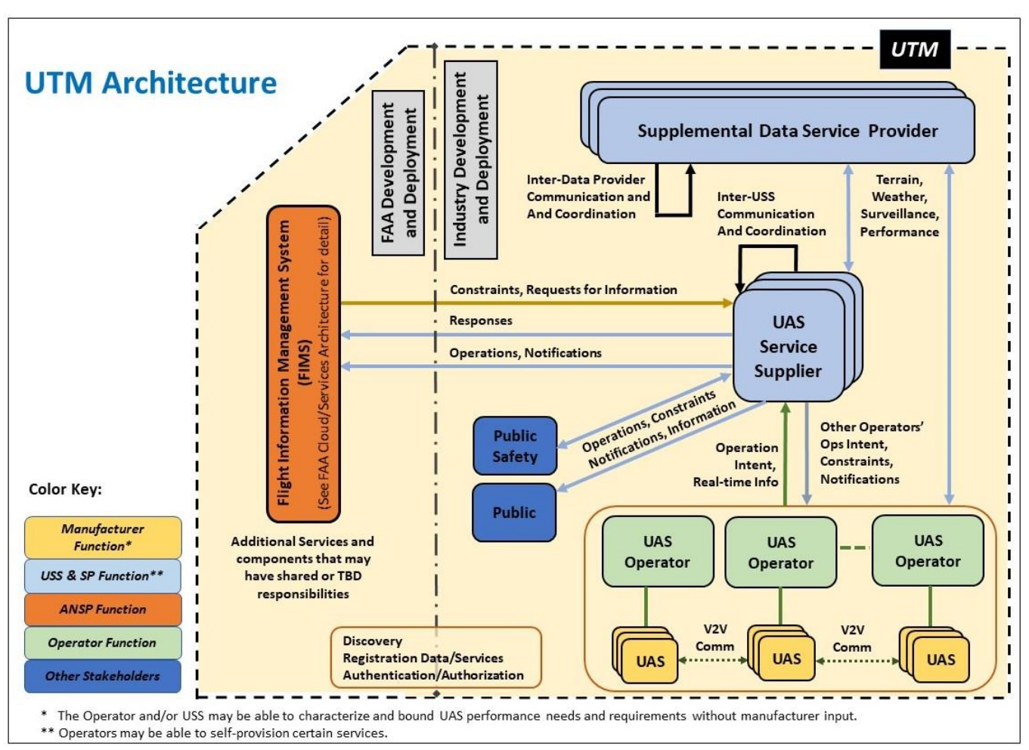}
    \caption{Current version of NASA UTM system \cite{NASAUTMConOps} }
    \label{UTM}
\end{figure}

As can be seen from Figure \ref{UTM}, the UTM system has two parts. The first part has the FAA as a central authority. The second part relates to development and deployment. While these parts are considered independent, they exchange different types of information. 
The FIMS is a central component of the UTM ecosystem. It resides in the FAA part of the system and connects unmanned service suppliers to the National Airspace (NAS) data center. It also enables airspace control, facilitates requests, and supports responses to emergency cases that impact the national airspace. Another primary agent in the UTM system is the operator that represents any entity responsible for operating a UAV in the airspace defined by the UTM system. On the other hand, a USS provides UAV operators with services such as separation, deconfliction, weather, flight planning, contingency management, and emergency information \cite{Unmanned2018NASA}. The UTM system enables the collaborative work of multiple USSs towards managing the low-altitude airspace without FAA intervention. Each USS is provided by a third party that manages the data flow and controls the processes of its participating drones. The use of multiple USSs enhances the system efficiency by distributing the management workload among different entities and avoids the situation of a single point of failure. Some of the data required to support the operation of unmanned services are provided to the USS by a separate agent called Supplemental Data Service Provider (SDSP). Other agents of the UTM system are the public and the public safety that includes the FAA, law enforcement, Department of Homeland Security, and other concerned governmental parties. For a detailed overview of the NASA UTM system and a comparison with other UTM initiatives, the reader is referred to \cite{shrestha2021survey}.  

\subsubsection{Remote ID}
RID is a technique for identifying a UAV by a ground agent or other airspace users. It aims at enhancing the safety, privacy, and security of drone operations in urban airspace. Especially for night flights, BVLOS flights, and operations over people, remote identification is regarded as an indispensable requirement. According to the FAA's final rule \cite{finalRule}, most drones are required to broadcast a remote ID while flying. RID messages comprise information about the drone ID, its location, altitude, velocity, take-off location, or the location of the control station, an emergency flag, and a timestamp . These messages can be sent using different communication techniques such as WiFi and Bluetooth \cite{ishihara2019remote}. RIDs are mainly allocated and managed by USSs. 

\subsection{Blockchain and smart contracts}
Blockchain is a distributed ledger technology (DLT) that has been deployed in a wide range of applications successfully. It enables independent entities to interact securely without the need for a trusted third party (TTP) \cite{wust2018you}. Smart contracts are computer algorithms that run on top of a blockchain leveraging its key features such as logs immutability, traceability, and auditability. Further, the deployment of smart contracts on the blockchain reveals various advantages. Particularly, smart contracts are autonomous, accurate, and secure by design. More importantly, the employment of smart contracts in the UTM system would reduce the time needed to obtain an authorization to fly. Another important advantage of smart contracts is the reduction of fraud. That is, once deployed, a smart contract becomes like a law that no one can bypass. We exploit these advantages to enforce the rules and regulations that control access to airspace.  

Although blockchain-enabled UAV applications are becoming more popular, only a few papers have focused on realizing a fully distributed UTM system. Rather, blockchain is usually used to allocate RIDs \cite{yazdinejad2020enabling} and record the communications within the UTM context \cite{allouch2021utm}. In contrast, our solution exploits the full potential of smart contracts by implementing a full UTM system that allows for allocating RIDs, planning and tracking missions, as well as enforcing airspace traffic rules and regulations.

\subsection{Mobile Crowdsensing }
Mobile Crowd-sensing (MCS), on the other hand, is an effective technology that exploits the capabilities of mobile devices to collect, forward, and process data \cite{feng2017survey}. MCS provides remarkable advantages over traditional sensing and monitoring methods as it does not require the physical installation of a wireless sensor network (WSN). Instead, MCS exploits human intelligence as well as the ubiquity and mobility of smart devices to provide low-cost sensing capabilities. Further, this technology is easily scalable to cover large and dynamic areas. We, hence, leverage these key features to design an effective scalable UAV monitoring system that seamlessly enforces the UTM regulations.
% It is worth noting that the FAA is currently developing a strategy to assign a session ID for each flight. The final rule on session ID is yet to be released. Therefore, we take a proactive move to design an end-to-end identification approach that considers both the remote ID as well as the session ID. 

\section{Related Work}\label{literature}
In response to the breakthrough of the blockchain technology, we observe a shift from centralized to decentralized systems. The synergy between blockchain and smart contracts offers various advantages including transparency, security, accuracy, efficiency, and trust, which are all vital for a sustainable management system. The advantages of decentralized systems for UAV operations were recognized and multiple related applications were proposed for supply chain \cite{fernandez2018uav}, surveillance \cite{islam2020bumar}, {situation supervision} \cite{islam2021blockchain}, {data acquisition} \cite{islam2019buav,islam2019bus}, coordinated UAV services \cite{sharma2017socializing}, and edge computing \cite{sharma2019neural}. Nonetheless, only a few papers attempted to design an end-to-end air traffic management system. In this section, we review these papers and highlight the novelty of our contribution. 
%Shrestha \textit{et al.} \cite{shrestha20216g} envisioned a 6G enabled UTM system that leverages the advantages of the 6G communication networks. In addition, the authors proposed an urban airspace segmentation and multi-layer airspace model which supports flight deconfliction. Nonetheless, the authors argue that the full implementation of this future perspective of the UTM system is associated with a multitude of challenges that need to be addressed.  
Rahman \textit{et al}. \cite{rahman} proposed a blockchain-based UTM system to ensure collision-free operations. Routes were planned in a way that avoids restricted areas such as private properties. Also, the system reduces the collision risk by minimizing the number of drones flying at the same height, which is specified as part of the mission plan. To enforce the drone to follow the specified route, the authors developed a smart contract to log drone movement and location information during the entire mission. If any observed attribute violates the specified flight route, a negative point is added to the drone's reputation. This solution, however, does not prevent uncooperative or malicious drones from logging fake data about their locations. Indeed, a more sophisticated monitoring mechanism is crucial to enforce rules and prevent such violations.

In the same context, Yazdinejad \textit{et al}. \cite{yazdinejad2020enabling} proposed a decentralized zone-based system for registering and authenticating drones. In the proposed architecture, the authors assign a ground-based agent that is in charge of managing the authentication process within a predefined perimeter. The availability of the authentication scheme is enhanced by allowing adjacent drone controllers to take over and substitute for a failing agent. However, this solution cannot be considered fully decentralized due to the reliance on managing agents through a trusted third party. Moreover, this work does not enforce mission rules rendering the solution prone to security threats by unlawful drones. 

Alternatively, Allouch \textit{et al.} \cite{allouch2021utm} proposed, implemented, and evaluated a permissioned blockchain to secure the UTM system. The system performs secure path planning and data sharing among participating drones. To deal with the limited resources on the UAV, the authors proposed to offload the computation to a cloud server while employing a decentralized off-chain storage system. Moreover, they excluded the participating UAVs from the peer-to-peer network and only allowed ground control stations to store a copy of the ledger. To evaluate their architecture, the authors implemented the solution on the Hyper-ledger Fabric platform and estimated both the delay and the resource consumption for transactions. The average latency of an invoke transaction on a network of 50 users was 454 ms. 
Despite exhibiting significant latency compared to existing communication networks, this work has shown promise for the application of blockchain-based UAV networks in real-time. 

Using state-of-the-art techniques to organize and control the airspace should be accompanied by mechanisms to enforce related rules and monitor the compliance of each user. None of the reviewed articles has focused on this aspect of the UTM system. Yet, many have explicitly highlighted the necessity of employing a monitoring and tracking mechanism to mitigate potential misuse. The recent concept of operation highlights the importance of ensuring accountability of operators and other actors in the UTM system \cite{NASAUTMConOps}. Nonetheless, this task is implicitly left to federal officers as per the recent regulations. This approach requires substantial effort and is prone to human errors.  Snead et al. \cite{snead2018establishing} highlighted multiple issues related to the capabilities of law enforcement and national security agencies in detecting, locating, and identifying unlawful drones. Moreover, this human-centric approach is non-scalable and not able to simultaneously monitor the expected air traffic volume.
Alternatively, the authors of \cite{ishihara2019remote} proposed an approach that enables the identification of participating and non-participating drones defining a communication protocol between the central authority, the USS, and the vehicle registration and model database. The approach is composed of eight negotiation steps between the aforementioned agents to retrieve registered information about the operator, the UAS properties, and the flight plan. However, this approach is fully dependent on law enforcement officers and thus not scalable. 

To the best of our knowledge, our work is the first to introduce a crowd-sensing approach to monitor the compliance of airspace users and thus incentivize the latter to act lawfully. We are also privileged to establish a fully decentralized automated regulatory framework that serves the objectives of the UTM system. Our solution utilizes existing matured technologies and concepts such as the UTM architecture and the remote ID. We develop a smart contract that can manage the end-to-end authorization process independently of any third parties. In addition, the utilization of the blockchain concept guarantees the accountability of all participating agents.

\section{System Overview}\label{overview}
\begin{figure*}
    \centering
    \includegraphics[width = \textwidth]{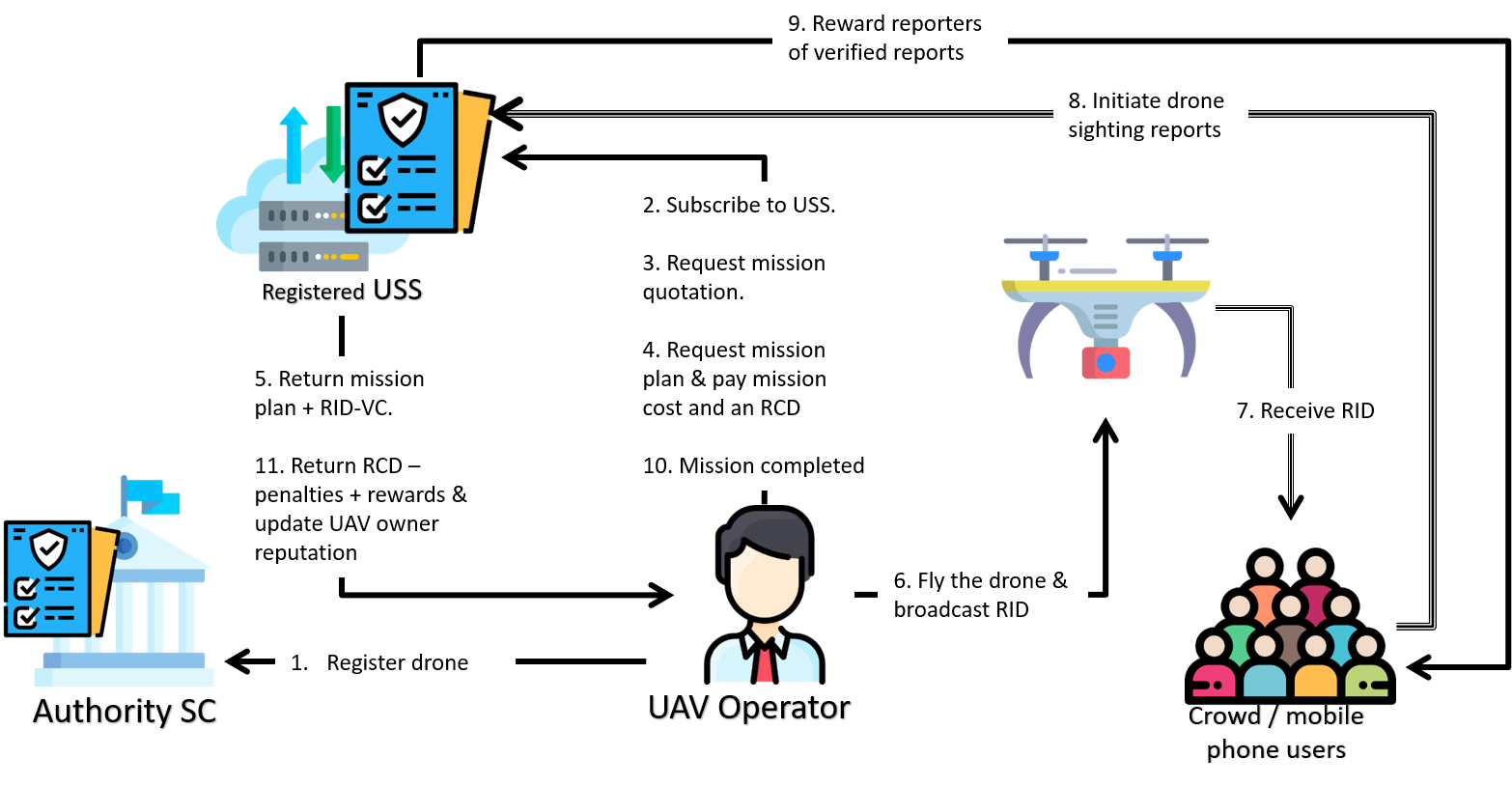}
    \caption{Overview of proposed UTM system.}
    \label{fig:overview}
\end{figure*}

The ultimate goal of the UTM ecosystem is to ensure high levels of privacy, safety, and security for airspace users and the public underneath the controlled airspace. We, therefore, take the research on the optimum UTM system one step forward by proposing a blockchain-based protocol that is designed to securely manage and control the access to the airspace by UAV operators. 
Figure \ref{fig:overview} illustrates an overview of the proposed system. In the first step, the UAV operator registers his drone by calling the registration function in the authority smart contract. As a result, the authority assigns an ID to the particular drone and adds the new drone information to its database. The database is only shared with registered UAV service suppliers (USSs). 

Then, the UAV operator subscribes to a USS to be able to schedule his missions. Although the current UTM architecture allows UAV operators to schedule their flights, it is believed that involving a USS in this task is vital for more reliable airspace management. Our model, therefore, assumes a subscription to a certified USS for flight planning. Indeed, the UAV operator is required to pay the USS subscription fees annually, quarterly, or monthly based on the selected subscription plan. 

Once subscribed, the UAV operator may have access to a multitude of services provided by the USS, including the ability to request a mission quotation. This request informs the UAV operator about the cost of his intended mission, which varies depending on the owner's reputation and the airspace congestion. Consequently, the UAV operator may request a mission plan and pay the estimated cost in addition to what we call Refundable Compliance Deposit (RCD) by calling the corresponding function in the USS smart contract. RCD refers to a fixed amount paid by the UAV operator to the USS smart contract when requesting a mission plan and returned back to the operator once the mission is over. Any penalties (fines) caused by UAV violations during the flight are deducted from the RCD before being returned to the UAV operator. The RCD can also be used to add rewards when the drone flies obeying the mission plan and rules.

The USS schedules the mission in accordance with the latest geofencing updates while ensuring route-deconfliction from concurrent UAV flights. In response to his request, the UAV operator gets the mission plan details along with a \emph{Remote ID Verification Code} (RID-VC) which needs to be broadcast along with remote ID during the flight. \textcolor{black}{The RID-VC is the hash of the plan details concatenated with a nonce.} The USS, keeps a record of the nonces to be able to verify the RID later. At this stage, the UAV shall be ready to start the mission. The UAV operator is committed to keeping broadcasting the RID-VC as well as the remote ID as per the latest FAA rules and regulations (drone serial number, location, velocity, timestamp, etc.). 

\begin{figure}
\centering
\includegraphics[width= \linewidth]{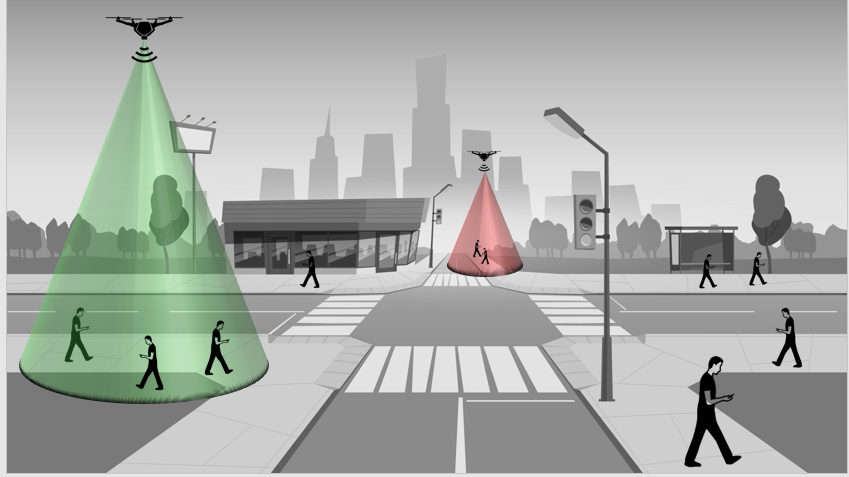}
\caption{Crowdsensing model }
\label{CS}
\end{figure}

\begin{figure*}
    \centering
    \includegraphics[width = \textwidth]{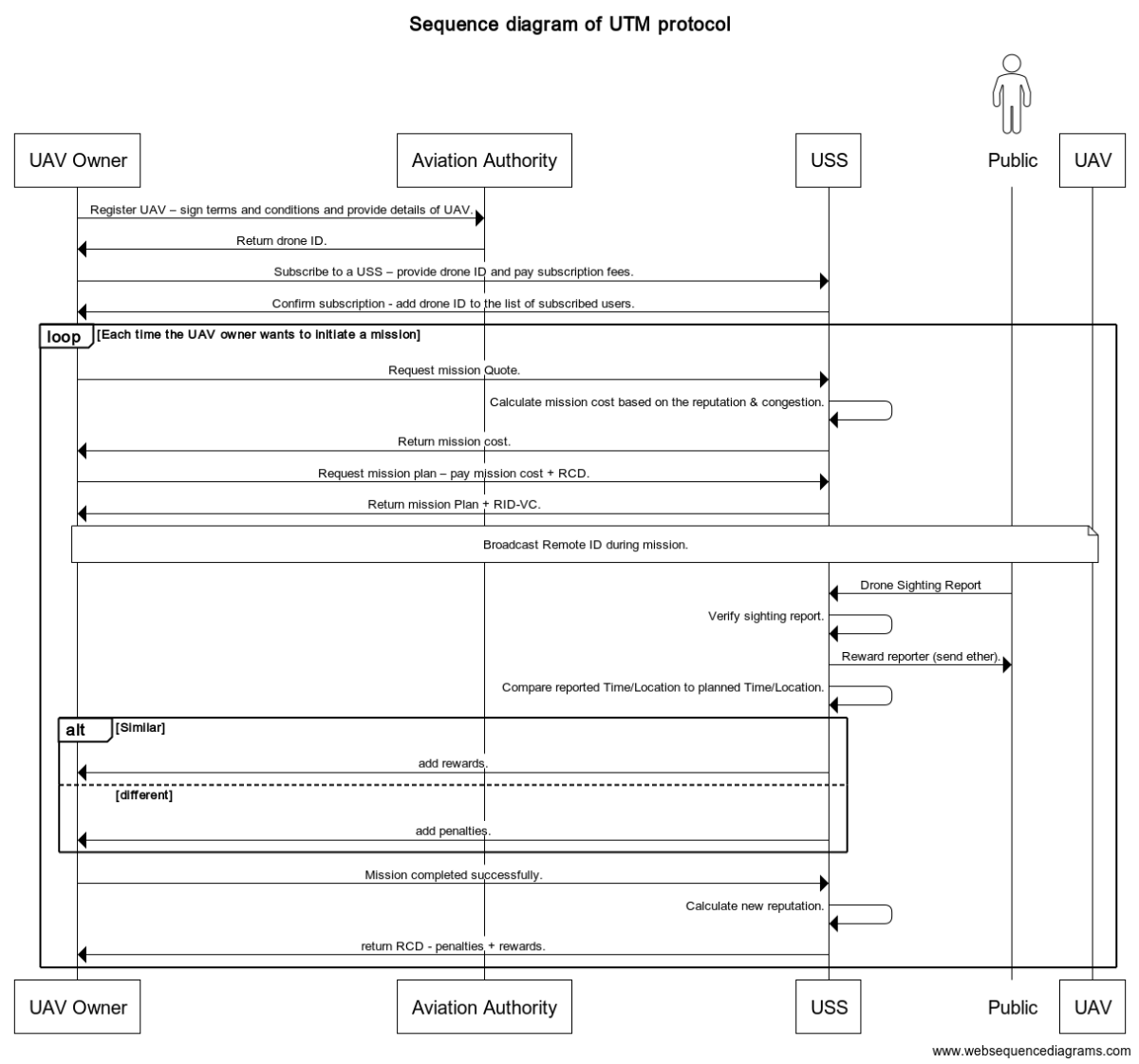}
    \caption{Sequence diagram of proposed UTM protocol.}
    \label{fig:my_label}
\end{figure*}

On the other hand, a mobile crowdsensing (MCS) model is exploited to monitor the intended mission and ensure that the UAV follows the assigned plan. Indeed, the employment of the MCS technology in this context improves the system coverage and saves the costs of installing dedicated monitoring systems. Smart devices have various communication modules such as WiFi that can be used to receive Remote ID information. We rely on the users of such devices to monitor drone behavior in the airspace by allowing the users to report the presence of a UAV at a specific time and zone by receiving and forwarding the RID broadcast as shown in Figure \ref{CS}. Public users are entitled to call the Sighting Report Function (SRF) on the UTM smart contract, which verifies the request and rewards the reporter in return. The reward can be a small amount of cryptocurrency or a voucher to encourage public contribution to rule enforcement. A voluntary MCS scheme can also be adopted. However,  we believe that the incentivized scheme is more effective in this context, because it encourages better participation and commitment among users. Indeed, multiple measures shall be put in place to prevent fake or dishonest reports. Such measures are out of the scope of this study. Interested readers are referred to \cite{handaja2020capp, truong2019trust} for a discussion and some examples. The received reports are compared with the original UAV mission plan to issue reward and penalty points accordingly.  
At the end, the UAV operator reports that the mission is completed so that the RCD can be refunded. The USS smart contract verifies the request and checks the rewards and penalties, which may be added to or subtracted from the RCD. Furthermore, the UAV operator's reputation is updated. This updated reputation shall be used in the calculation of the next mission cost. Figure \ref{fig:my_label} illustrates the sequence of communication between different parties to initiate a mission.

\section{{System Design}}\label{implementation}
\begin{figure}
    \centering
    \includegraphics[width = \linewidth]{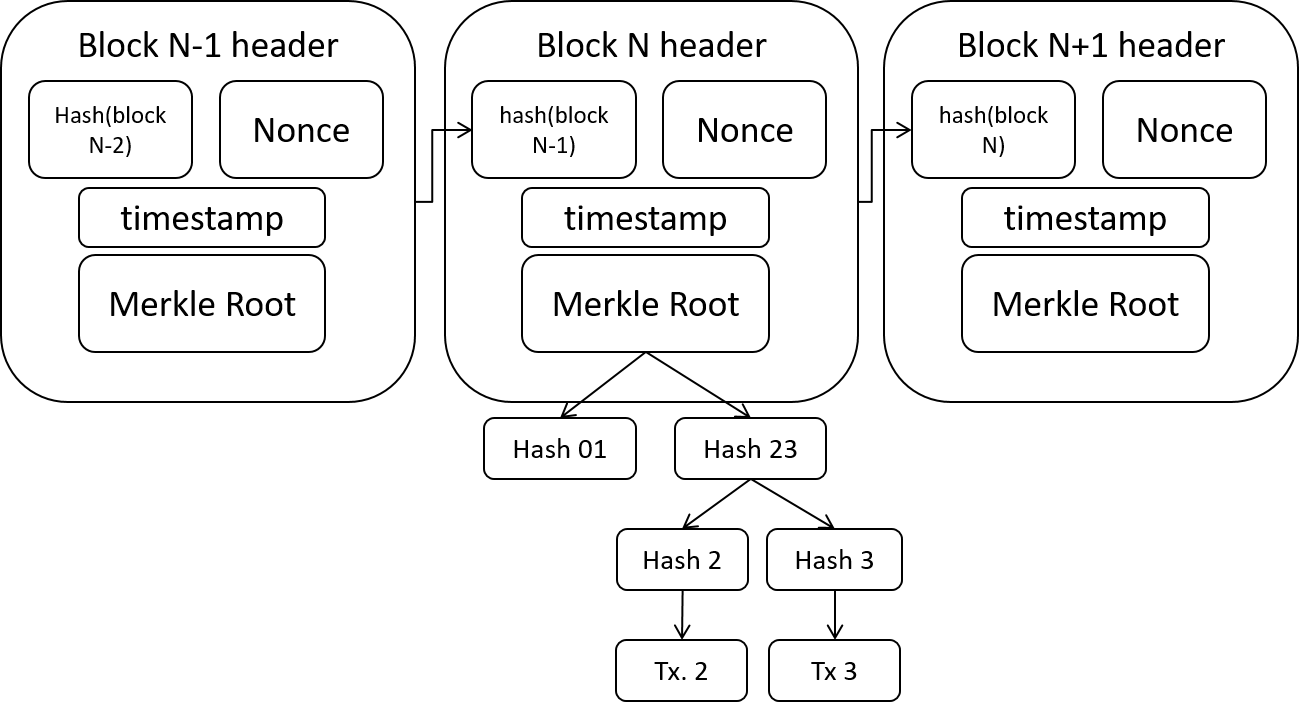}
    \caption{{Block structure in the proposed blockchain.}}
    \label{block}
\end{figure}

{In this section, we describe the blockchain architecture and the outline of the designed smart contracts. To be able to serve the public users, we opted to design a public blockchain that operates in a permissioned mode. Particularly, we implemented our solution on top of the Ethereum blockchain where users can join at their convenience. Yet, access control modifiers are utilized to manage the privileges and authorities of users. Ethereum is a public blockchain that stores information and smart contracts on a public distributed ledger. In our case, information about the drone, its operator, and its missions are stored on the Ethereum blockchain. Each block contains multiple transactions that store data and states of the smart contracts. The block structure is illustrated in Figure }\ref{block}. 
{Currently, the Ethereum blockchain uses the Proof-of-Work (PoW) consensus mechanism to validate and add blocks to its main blockchain. In particular, miner nodes compete to verify the validity of the information and transactions executed by the smart contracts before solving a mathematical challenge. Basically, this challenge requires producing the block nonce through a trial and error strategy. Upon the successful production of the block nonce, the miner broadcasts his block to other nodes in the network to verify it and add it to the main chain. The miner receives the transaction fee as a reward in return. Illegitimate data modification is not possible, and any change in the state of the contract requires the authorized entity to invoke a certain function which again creates a new transaction. This way guarantees the integrity of the data on the blockchain. Although using a public blockchain such as Ethereum is more expensive than establishing a private blockchain where less-expensive consensus mechanisms (e.g. Proof-of-Authority) are adopted, it is argued that private blockchains are prone to centralization issues, security breaches, and service discontinuity }\cite{haque2020blockchain}.    

\subsection{Register Drone} 
\begin{algorithm}
\SetAlgoLined
\KwResult {Add drone information to the registeredDrone structured array}
 \textbf{Inputs:} caller, droneSerial, ownerId, signTAC\;
  \textbf{Output:} {droneId}\;
 \eIf{droneSerial $\in$ registeredSerial[]\;
 }{
 revert: 'Drone already registered'\;
 }{
  \eIf{signTAC}{
     set Drone(droneSerial, ownerId, rewards, penalties, hasActivePlan, msg.sender)\;
     push registeredDrones[i]= Drone  \;
     set droneId = i\;
   %  emit NewDrone[id]\;
     return droneId\;
   }{
    revert: 'Please accept terms and conditions'\;
    
    }
 }
 \caption{Register Drone}
\end{algorithm}

Upon purchase and before take-off, a UAV operator is requested to register her drone in the Aviation Authority. This is accomplished by calling the registerDrone function in the authority Smart Contract (SC). This function takes the drone serial number and the operator's national ID number as inputs. Also, the operator is requested to read and sign the terms and conditions for using the public airspace. This is presented in the signTAC Boolean input. The smart contract function checks if the drone's serial number already exists in the list of registered drones. After verifying the signature of terms and conditions, the drone and operator information is added to the registeredDrones array. Finally, the function returns the new drone ID, which is the index of the drone in the registeredDrones array. This ID will later be used to identify the drone in all communications. 

\subsection{Subscribe to USS}

\begin{algorithm}
\SetAlgoLined
\KwResult{ Add registered drone to USS subscription list for a certain fee.\; }
 \textbf{Inputs:} caller, droneId\;
 \textbf{Outputs:} {Update SubscribedDrone array}\;
 \eIf{caller != registeredDrone[droneId].ownerEA\
 }{
 revert: 'Not the owner of the registered drone'\;
 }{
  \eIf{SubscribedDrone[droneId] == 0}{
     revert: 'Drone is already subscribed'\;
   }{
    \eIf{msg.value == SubscriptionFee\
    }{
    SubscribedDrone[droneId]= caller\;
    }{
    revert: 'Please make sure to pay the subscription fee'\;
    }
    
    }
 }
 \caption{Subscribe to USS}
\end{algorithm}

To benefit from USS services, a drone owner needs to subscribe to an authorized USS by calling the SubscribeToUSS function in the particular USS SC and paying the subscription fee. As shown in algorithm 2, this function takes one argument (droneID) and checks if the caller of the function owns this drone. Then, it checks if the drone is already on the subscribed drones' list. Finally, it checks the amount of cryptocurrency sent to the contract to make sure that it is equal to the required subscription fee. If all the checks are successful, the droneID is added to the subscribed drones' list. 
\subsection{Request mission quotation}

\begin{algorithm}
\SetAlgoLined
\KwResult{ Returns mission dynamic cost }
 \textbf{Inputs:} callerEA, droneId\;
  \textbf{{Output:}} $F_D$\;

 \eIf{callerEA != registeredDrone[droneId].ownerEA\
 }{
 revert: 'Not the owner of a registered drone'\;
 }{
  \eIf{SubscribedDrone[droneId]==callerEA}{
  return ($F_D = kd+c+a$)\;

   }{
         revert: 'Drone is not subscribed'\;

    }
 }
 \caption{Request Mission Qoute}
\end{algorithm}
Before each flight, the drone owner is requested to obtain a mission plan and a RID. The Request Mission Plan (RMP) function assists the user in getting this information.  But before requesting the mission plan, the user is asked to obtain a mission quote, which is the fee required to reserve the airspace and schedule a plan for his mission. Inspired by the discussion in \cite{dasu2018geofences}, this fee is designed to be dynamic to help reduce congestion in peak hours. The dynamic fee for the mission is calculated using the following expression: 
\begin{equation}
  F_D = kd+c+a;  
\end{equation}

where $F_D$ is the dynamic fee, $k$ is the cost scaling factor that depends on the reputation of the UAV operator (as will be shown in equation \ref{eq-3}), $d$ is the original cost of the mission when no congestion is assumed, $c$ is the refundable compliance deposit (RCD), and $a$ is a variable amount of cryptocurrency that is proportional to the congestion rate. 

Once the function is called by the UAV operator, the calculation is made and the dynamic fee value is returned to the caller.

\subsection{Request mission plan}

\begin{algorithm}
\SetAlgoLined
\KwResult{Return mission plan, RID-VC and calculate and receive the dynamic fee \; } 
     \textbf{Inputs:} caller, droneId, sourceLocation, destinationLocation, departureDate, departureTime  \;
     \textbf{Outputs:}{ Plan} \;
 \eIf{caller $\in$ SubscribedDrone[] }{
   \eIf{registeredDrone[droneId].hasActivePlan}{
     revert: 'There is already an active plan for this drone'\;
   }{
   $F_D = kd+c+a$\;
   
    \eIf{msg.value $\ge$ $F_D$   }{
    RIDVC = Hash(nonce $\parallel$ caller $\parallel$ sourceLocation $\parallel$ destinationLocation $\parallel$ departureDate $\parallel$ departureTime)\;
    set Plan(coordinates, departureTime, departureDate, altitude, RIDVC)\;
    push missionPlan[i]=Plan\;
    registeredDrone[droneId].hasActivePlan = true\; 

    return Plan\;
    
    }{
    revert: 'Please make sure to pay the mission plan fee'\;
    }
    
    }
 }{
    revert: 'Not subscribed to a USS'\;
 }
 \caption{Request mission plan}
\end{algorithm}
 To request the mission plan and obtain an RID-VC, the requestMissionPlan function on the USS-SC shall be called by the drone owner. The function permits the user to input his Drone ID, the source and destination locations, and the mission date and time. The function checks if the owner has a valid subscription to one USS and if the drone does not already have an active mission plan. Then, the contract checks the amount of cryptocurrency paid to the contract to ensure that the owner has paid at least the required dynamic fee.  The RID-VC is calculated by concatenating a nonce with the input information entered by the user.  This RID-VC is used to ensure that the user does not deny the mission plan, and that the plan is scheduled by an authorized USS. Finally, the mission plan is scheduled (by some scheduling algorithms \cite{razzaq2018three,akgunduz2017deconflicted,sacharny2020faa}) and returned with the RID-VC to the user. 
 
 During the mission, the UAV operator is responsible to broadcast his RID as per the latest FAA concept of operation \cite{NASAUTMConOps}. The transmitted remote ID shall take the following form: 
 \begin{equation} \label{RI_D}
 RID = RID_{FAA} || RID_{VC};
 \end{equation}
 where $RID_{FAA}$ is the concatenation of the information that the FAA requires each drone to broadcast. Currently, this term shall look as follows: 
  \begin{equation} 
 RID_{FAA} = T_s || L_d || L_{CS} || A || v;
 \end{equation}
 where $T_s$ is the timestamp of the broadcast $L_d$ is the current location of the drone, $L_{CS}$ is the location of the control station, $A$ is the altitude of the drone, and $v$ is the velocity. This structure of the message ensures the integrity of its information. Besides, the cryptographic structure of the RID message agrees with the FAA requirement \cite{NASAUTMConOps} which states that "the FAA may require the RID to be cryptographically protected by an authentication message, ensuring the authentication, non-repudiation, and integrity". 
 
\subsection{Report Drone}

\begin{algorithm}
\SetAlgoLined
%\KwResult{ } 
 \textbf{Inputs:} caller, droneId, RID, sightingLocation, sightingTime  \;
 \textbf{Outputs:} {Verify sighting report, pay reporter, add rewards or penalties } \;
 \eIf{caller == registeredDrone[droneId].ownerAdd}{
 revert: 'Owner of drone cannot report it!'
 }{
 \eIf{reportsPerDroneByCaller $\ge$ 1}{
 revert: 'not allowed to report same drone more than once '\;}{
 reportsPerDroneByCaller++\;
 
 \eIf{verifyRID()}{
 reward caller\;
 emit DroneSighted(droneId, SightingLocation)\;
 sightingTime = now\;
 \eIf{sightingLocation == missionPlan[droneId].coordinates $ \& $ sightingTime == missionPlan[droneId].time
 }{
 registeredDrones[droneId].rewards++\;
}{
registeredDrones[droneId].penalties++\;
}
}{
revert: 'Invalid report'\;
}
 }}
\caption{Report Drone }
\end{algorithm}

The drone needs to keep broadcasting its RID while in the mission. The public is incentivized to report drones by calling the ReportDrone function. This function takes four arguments: the drone ID, broadcasted RID, sighting location, and sighting time. The drone owner is not permitted to report his drone. Also, the same reporter is not permitted to report the same drone more than once, to avoid unfair reports. Thus, the function checks to ensure these conditions are met. Then the function verifies the RID-VC with the given drone ID. If the verification is successful, the reporter is rewarded. Then, the contract checks if the sighting information is consistent with the plan of the drone. If yes, a reward is given to the drone, else the drone is penalized.

\begin{algorithm}
\SetAlgoLined
%\KwResult{End mission, settle payment,  and delete plan  \; } 
 \textbf{Inputs:} caller, droneId, RID-VC\;
 \textbf{Outputs:} {End mission, settle payment,  and delete plan } \;
 \eIf{caller == droneOwner}{
 
 \eIf{hasActivePlan}{
 
 caller.transfer(RCD - penalties + rewards)\;
 calculate R; 
 reset rewards \;
 reset penalties\;
 emit missionComplete(RID-VC)\;
 hasActivePlan = false\;
 delete(missionPlan[droneId]);\
 
 }{revert: 'No active plan'\;}}{revert: 'Not owner of drone'}
\caption{Report Mission Completion}
\end{algorithm}

\subsection{Report Mission Completion}
\begin{figure}
    \centering
    \includegraphics[width=\linewidth]{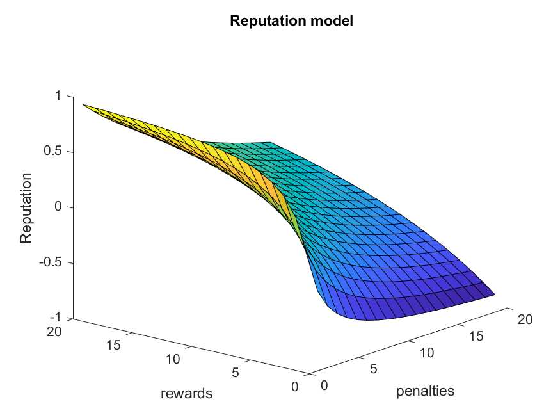}
    \caption{Expected value of the reputation. }
    \label{reputation}
\end{figure}

    After landing at the destination location, the owner of the drone needs to call the missionCompleted function to receive back the RCD fee paid earlier. However, in case the UAV was reported during the mission, then the rewards will be added to the returned RCD, whereas penalties are deducted from the same. Further, before resetting the rewards and penalties, they are used to calculate the reputation of the drone owner, which in turn affects the future mission plan fees. For simplicity, we use the Beta reputation system \cite{josang2002beta} which uses a modified expected value of the Beta distribution to estimate the reputation of a user. The reputation is calculated as follows:
\begin{equation}
    R = \frac{r-p}{r+p+2};
    \end{equation}
    
    where $R$ is the UAV operator's new reputation, $r$ is the number of reward points, and $p$ is the number of penalty points during the mission. Other more sophisticated reputation schemes \cite{putra2021trust} may also be employed to improve the detection and elimination of malicious users. However, the selection of the reputation mechanism is beyond the scope of this article. Figure \ref{reputation} shows the expected value of the reputation at different combinations of rewards and penalties. Consequently, the $k$ parameter which is used to calculate the next mission plan fee is defined as: 
    \begin{equation}\label{eq-3}
    k = (1-\frac{R+1}{2})a+ k_{prev}(1-a);
\end{equation}
where $0<a<1$ is a tunable parameter that defines how the new reputation affects the next mission plan cost, and $k_{prev}$ is the last $k$ value that the UAV operator used to have. Increasing the value of $a$ makes the model more sensitive to changes in reputation while decreasing $a$ yields a more stable cost function. Note that the value of range of $R$ is (-1,1). We, thus, shift the value by 1 and scale it by a factor of 0.5 to change the range to (0,1) which can be more easily interpreted.

\section {Testing and Validation }\label{test}
{To test and verify the proposed protocol, our algorithms were implemented on the Remix IDE \cite{remix} which is an online platform that is used to write, execute, debug, and test solidity-based smart contracts before implementing them on the real blockchain.} In this section, we test and verify the operation of the access control modifiers as well as the output of each function. Each function is thus tested multiple times using different Ethereum Addresses (EA) to ensure that only designated users can call the specific function. In our case, each UTM agent has a special EA including the UAV operator, the public reporter, the USS SC, and the SC owner (authority). The EA assigned to each agent during our simulation is provided in Table \ref{EA}. In this section, we describe the successful execution of each function and provide snapshots of the resulting transaction. 

\begin{table}[]
\caption{Ethereum addresses used to simulate the different transactions.}
\label{EA}
\begin{tabular}{|l|l|}
\hline
\textbf{Agent} & \textbf{Ethereum   Address} \\ \hline
\textbf{UAV operator} &\footnotesize{0xCA35b7d915458EF540aDe6068dFe2F44E8fa733c}\\ \hline
\textbf{Public reporter} &\footnotesize{0x17F6AD8Ef982297579C203069C1DbfFE4348c372}\\ \hline
\textbf{AuthoritySC} &\footnotesize{0x0DCd2F752394c41875e259e00bb44fd505297caF}\\ \hline
\textbf{USS\_SC} &\footnotesize{0xB87213121FB89CbD8B877Cb1Bb3FF84dD2869cfA}\\ \hline
\end{tabular}
\end{table}

\subsection{Register Drone}
\begin{figure}
    \centering
    \includegraphics[width=\linewidth]{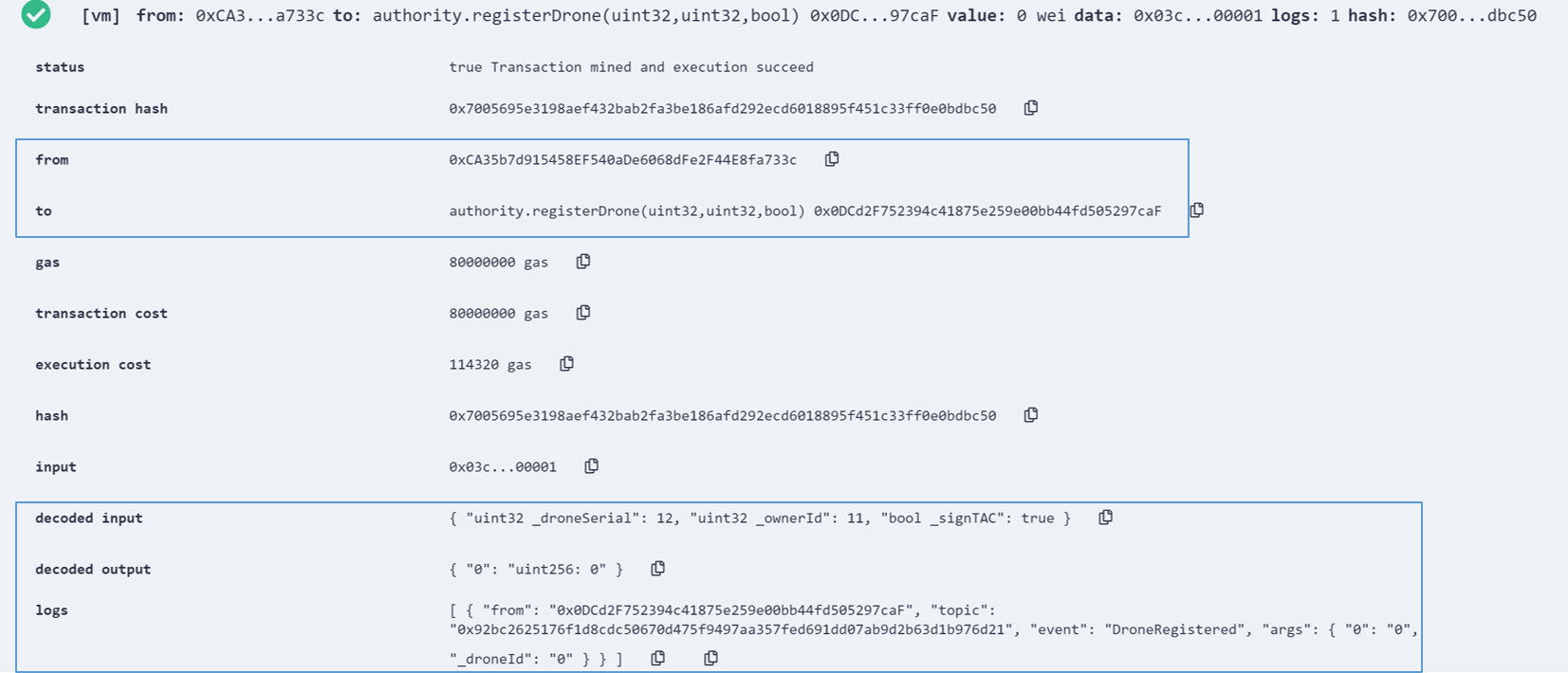}
    \caption{Sample transaction for executing the Register Drone function. }
    \label{regsiterDrone}
\end{figure}

This function is part of the authority smart contract. It allows UAV operators to register their drones under their names to be able to use the rest of the NAS services. We tested this function using the UAV operator EA. As shown in Figure \ref{regsiterDrone}, the function is successfully called by the UAV operator as can be noticed in the “from” and “to” fields. The “decoded input” field shows the arbitrary drone serial number and owner ID that was used in this example. Also, it shows that the owner agreed to the terms and conditions by setting the signTAC input to true.  Finally, the “decoded output” field shows the drone ID which is returned to the user to be used in the rest of the transactions. 

\subsection{Subscribe to USS}
\begin{figure}
    \centering
    \includegraphics[width=\linewidth]{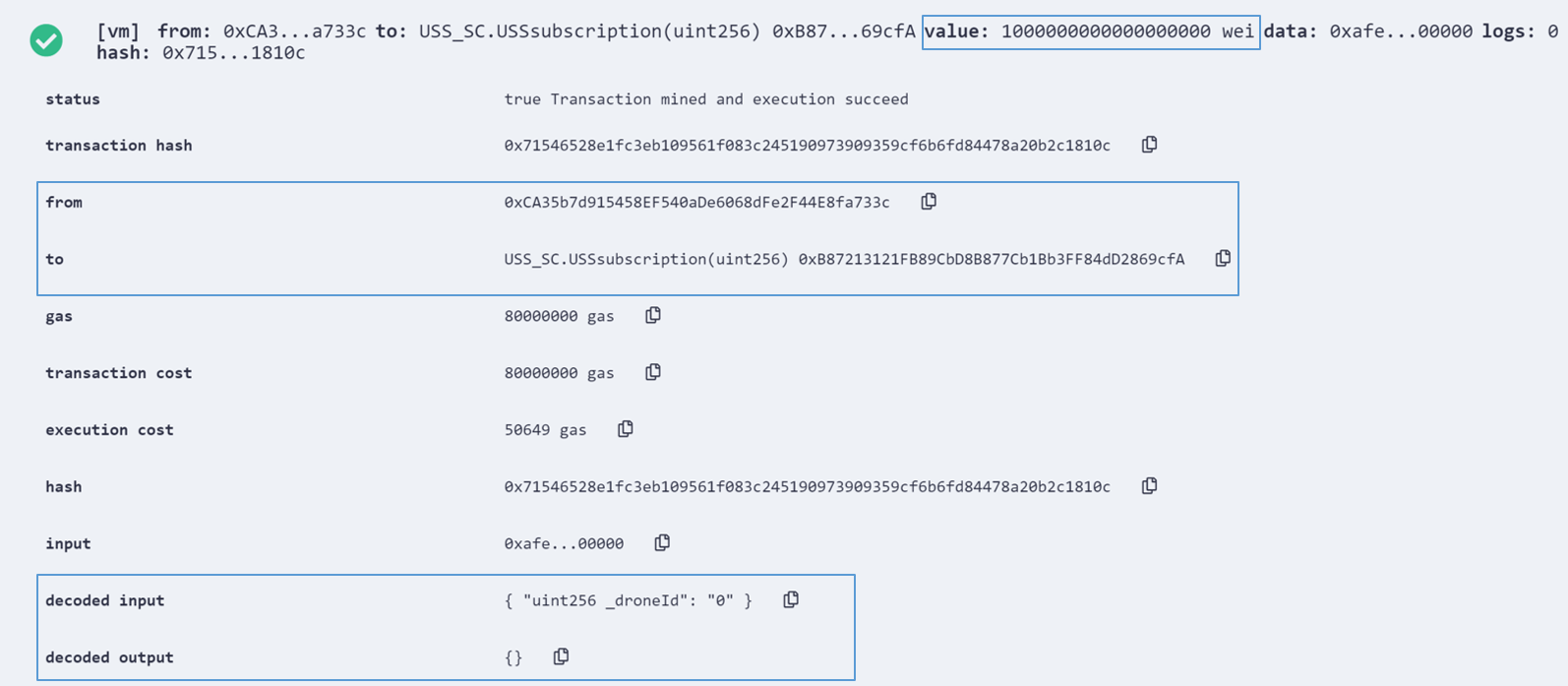}
    \caption{Sample transaction for executing the Subscribe to USS function. }
    \label{Subs}
\end{figure}

This function allows registered drones to subscribe to a USS to benefit from its services such as path planning, deconfliction, weather forecasting, emergency information, etc. The UAV operator needs to i) call this function from his address which was used to register the drone (see the “from” field in figure \ref{Subs}), ii) input the drone ID which he received when registering his drone (see “decoded input” field), and iii) pay the subscription fee as shown in the “value” field. Note that the subscription fee is set to 1 ether for demonstration purposes only. 

\subsection{Request Mission Quote}
\begin{figure}
    \centering
    \includegraphics[width=\linewidth]{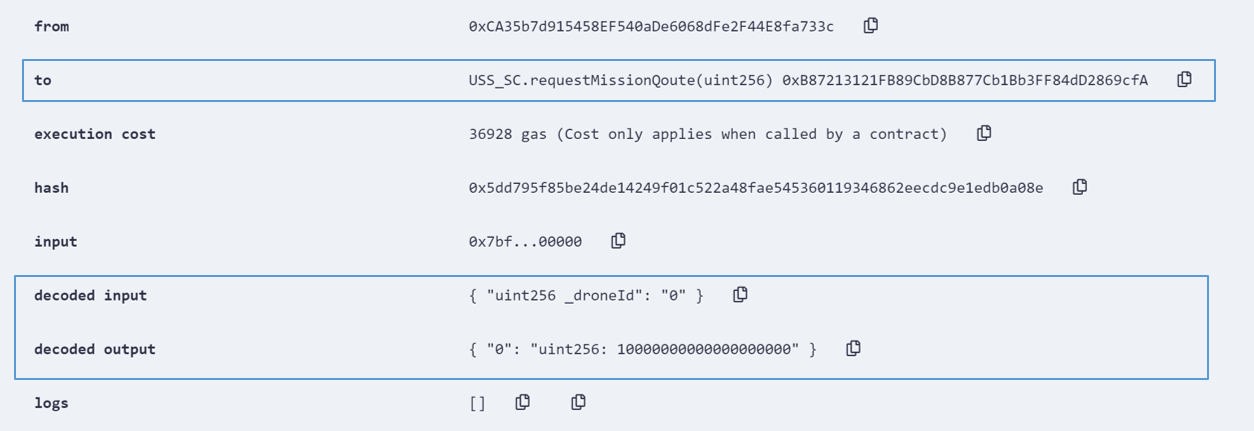}
    \caption{Sample transaction for executing the Request Mission Quote function. }
    \label{requestmissionQuote}
\end{figure}

This function is a view function that does not cost any gas. The only purpose of this function is to inform the user about the cost of his next mission plan. The UAV operator simply calls the function, inputs the drone ID, and receives the next mission cost. These details are shown in the “from”, “decoded input’, and “decoded output” fields in Figure \ref{requestmissionQuote}, respectively. 

\subsection{Request Mission Plan}
\begin{figure}
    \centering
    \includegraphics[width=\linewidth]{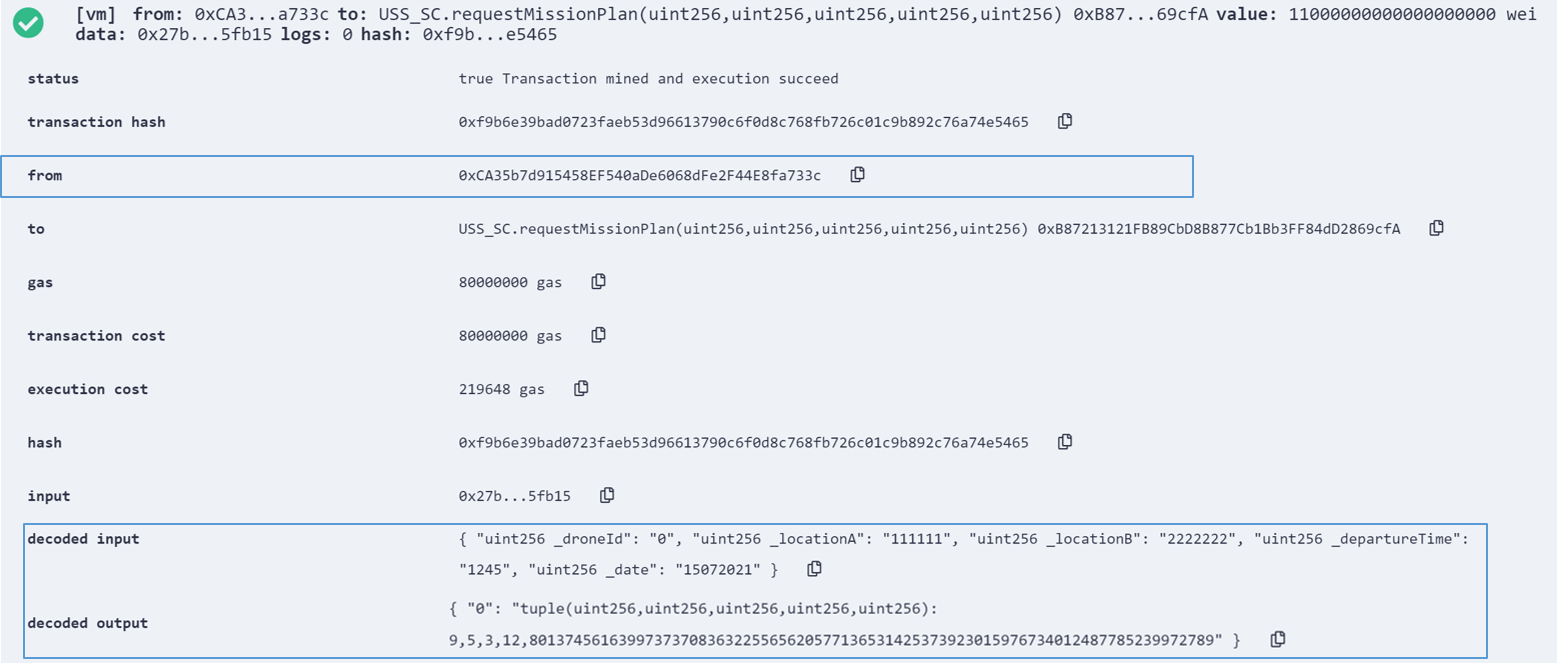}
    \caption{Sample transaction for executing the Request Mission Plan function. }
    \label{Plan}
\end{figure}

A UAV operator willing to fly his drone from point A to point B shall call the request mission plan function which resides in the USS-SC that he is subscribed to. The UAV operator needs to input his drone ID, the coordinates of the take-off location in Degree/Minute/Seconds (DMS) format, the coordinates of the destination, the departure time in the form (hhmm), and the date of the requested mission in the form (ddmmyyyy). These inputs can be viewed in the “decoded input” field, while the UAV operator address is shown in the “from” field (Figure \ref{Plan}). Finally, the value of the transaction reveals the amount paid by the owner to get the flight plan and the mission RID-VC (see the “decoded output” field).

\subsection{Report Drone}
\begin{figure}
    \centering
    \includegraphics[width=\linewidth]{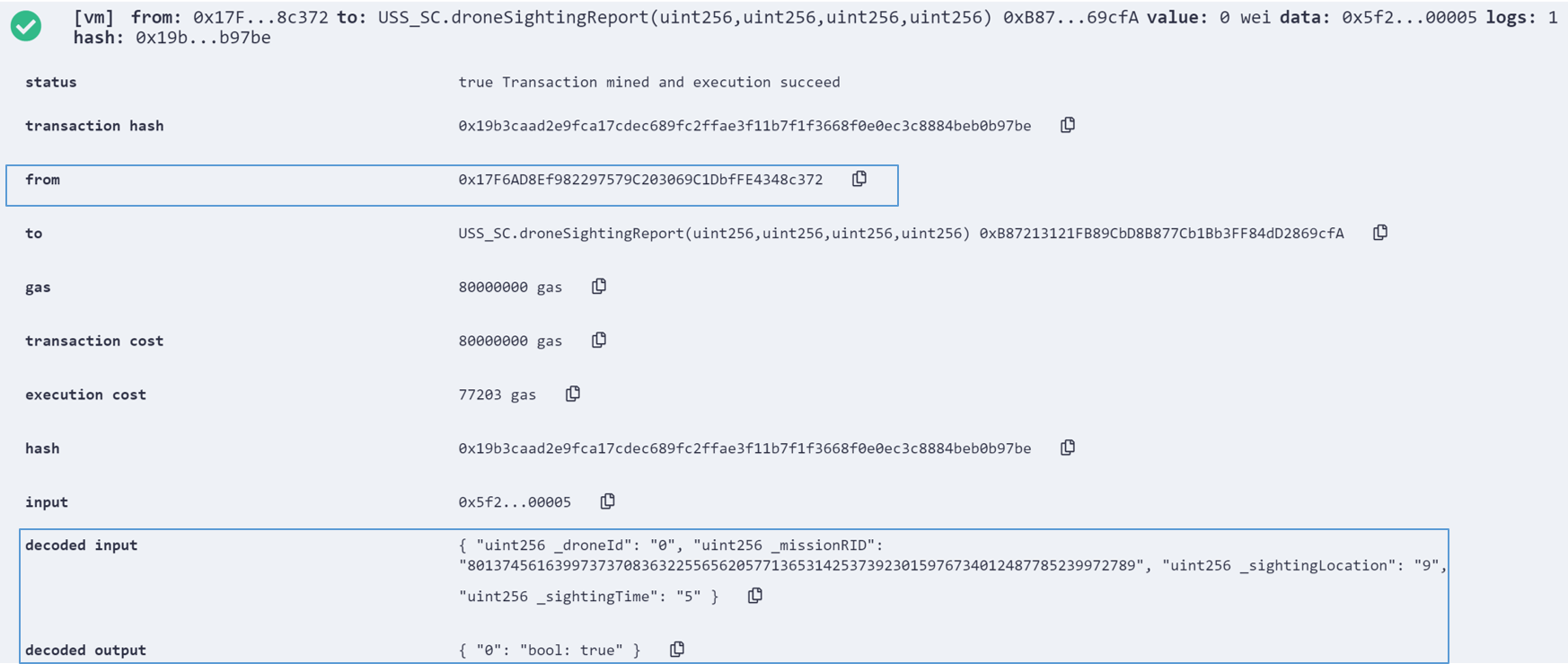}
    \caption{Sample transaction for executing the Report Drone function. In this case the UAV operator is rewarded.  }
    \label{Sighting}
\end{figure}

In this example, we call the ReportDrone function from a public reporter EA. The address of the caller is shown in the “from” field. The public user inputs four parameters to the function which are shown in the “decoded input” field. The first two parameters are the ones received by the user from the sighted drone. The other two are the user’s current location and time at which he saw the drone. The function tries to match the drone RID with the caller's location and time. In case they match, the UAV operator receives a reward. Otherwise, he receives a penalty point. The caller of the function is also rewarded in case the RID he sent is found to be valid. This is done by verifying the RID-VC. \textcolor{black}{Figure} \ref{Sighting} shows an example of a rewarded UAV operator. Note that the function throws an error if called by the UAV operator. Also, the same EA cannot call the function with the same drone ID more than once. These restrictions are placed to prevent greedy users from getting the reward without contributing important information.

\subsection{Report Mission Completion}
\begin{figure}
    \centering
    \includegraphics[width=\linewidth]{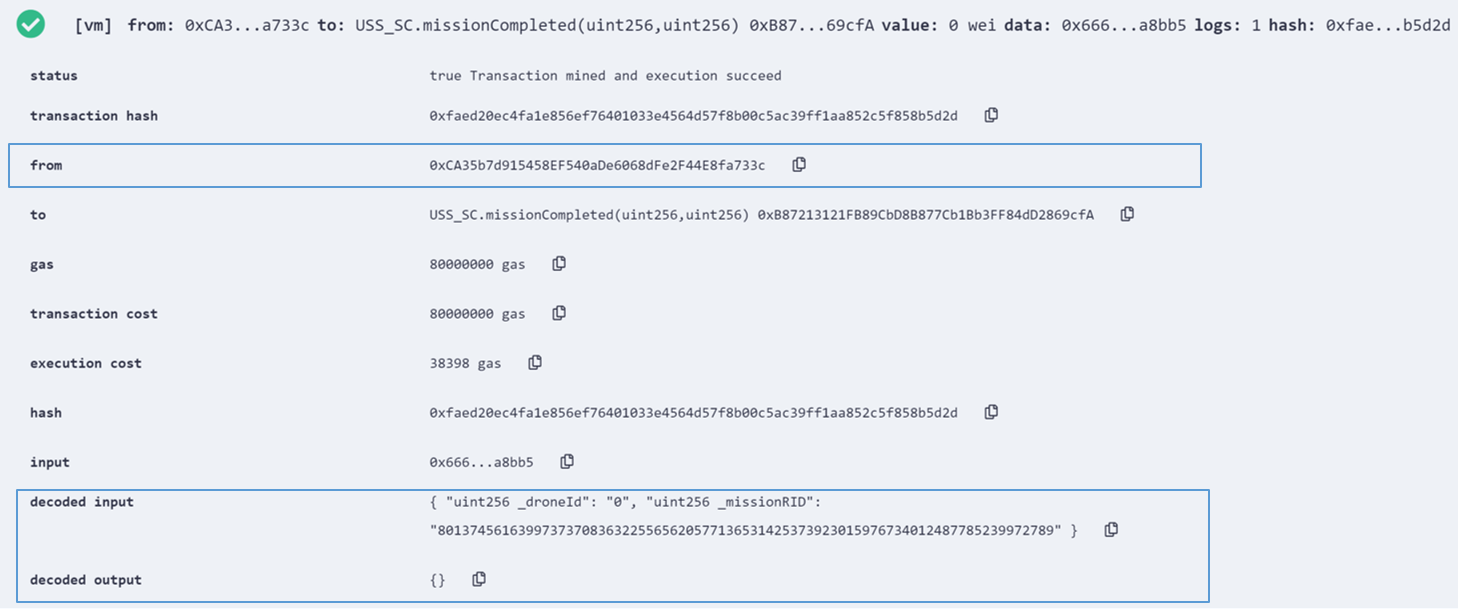}
    \caption{Sample transaction for executing the Report Mission Completion function.  }
    \label{completemission}
\end{figure}

After arriving at the destination, the UAV operator is responsible to  declare the end of his mission by calling the Report Mission Completion function. This function takes two arguments, the drone ID and the mission RID. Indeed, the function updates the reputation of the drone and deletes the mission from the list of active missions to allow the owner to request a new mission. It also returns the RCD to the caller after deducting/adding the penalties/rewards.  Figure \ref{completemission} shows the details of the simulated transaction.

\section{{Performance Analysis}}\label{Performance}

\subsection{{Cost Analysis}}

\begin{table}[]
\caption{Gas cost of smart contract functions }
\label{gas}
\begin{tabular}{|l|l|l|}
\hline
Function & Execution gas & Cost in USD \\ \hline
Register Drone & 114320 & 2.622 \\ \hline
Subscribe to USS & 50649 & 1.161 \\ \hline
Request Mission Quote & 0 & 0 \\ \hline
Request Mission Plan & 219648 & 5.039 \\ \hline
Report Drone & 77203 & 1.771 \\ \hline
Report Mission Completion & 38398 & 0.8809 \\ \hline
\end{tabular}
\end{table}

{In this section, we estimate the cost of executing each function in the proposed smart contract. In particular, we evaluate the cost of the Ethereum gas required to call each function on the blockchain using the REMIX IDE} \cite{remix}. Table \ref{gas} shows the execution gas and the corresponding cost in USD calculated using the average gas price (10 Gwei) retrieved from the Ethereum Gas Station \cite{EGS} on the $27^{th}$ of July 2021. On the same date, the price of one ether reached 2294.17 USD. As discussed above, the "request mission quote" function is a view function that does not change any state variables and thus does not incur any gas cost. The cost of the mission function is the lowest because it only changes a few state variables on the blockchain. In contrast, requesting a mission plan incurs the highest gas cost due to the complexity of the related function.  It is worth noting that the current gas and Ether prices are much higher than the average prices before December 2020. Such severe fluctuations in gas prices render any cost analysis less helpful. One solution is to use a private blockchain where only trusted miners are permitted. The mining cost would then be either predefined as a constant value or set to zero \cite{private_blockchain}. Indeed, the choice of such a solution would introduce a trade-off between decentralization and cost. The optimization of this tradeoff will be part of our future work.

The proposed MCS technique cancels the costs associated with the installation of UAV detection and tracking technologies. The availability and mobility of smart-device users allow for covering larger geographical areas without the need for expensive investments in such infrastructures. The proposed MCS technique can be seen as a pay-as-you-use service which can be an attractive model in many cases. Although a voluntary MCS scheme is conceivable, we believe that an incentivized scheme provides a more robust monitoring mechanism. In this case, the reporter incentives can be paid by the government or the USSs as they profit from the subscription fees and penalties deducted from the RCDs. Sophisticated cost models that consider such aspects will be part of our future work. 

\subsection{Comparison with related work}
In section \ref{literature},{ we reviewed the literature on blockchain-based UAV networks. In light of our review, we present a comparison between our proposed system and the systems presented in the literature. A summary of this comparison is provided in Table} \ref{comparison}.

\begin{table}[]
\caption{{Comparison between our work and related work in the literature.} }
\label{comparison}
\begin{tabularx}\columnwidth{|X|X|X|X|X|}

\hline
 & \textbf{Our solution} & \textbf{Rahman \cite{rahman}} & \textbf{Yazdinejad \cite{yazdinejad2020enabling}} & \textbf{Allouch \cite{allouch2021utm}} \\ \hline
\textbf{Blockchain platform} & Ethereum & Ethereum & NS3 & Hyberledger  Fabric \\
\hline
\textbf{Mode of  operation} & Permissioned public & Private & Public & Permissioned \\
\hline
\textbf{Planning missions} & Yes & Yes & No & Yes \\
\hline
\textbf{Policy enforcement} &  MCS & Wireless networks (cellular)  &  Distributed servers (cluster heads) & No \\
\hline
\textbf{Reputation system} & Yes & No & No & No \\
\hline
\textbf{Dynamic pricing } & Yes & No & No & No \\ \hline
\end{tabularx}
\end{table}

{Perhaps, the work of Rahman et al. }\cite{rahman} {is the most relevant to our work. They implemented a private blockchain to provide mission planning and policy enforcement. Yet, their policy enforcement mechanism relies on real-time logging of UAV coordinates via wireless networks such as cellular networks. Not only this solution is energy demanding, but also the coordinates can be easily manipulated by the malicious UAV operators to avoid being tracked and penalized. On the other hand, the MCS solution provides a more effective mechanism to track and penalize such uncooperative operators, while not imposing any processing overhead on the drone. Similarly, Yazdinejad et al. }\cite{yazdinejad2020enabling} {simulated a public blockchain using the NS3 network simulator. They aimed at enforcing identification and geofencing rules by employing a set of servers distributed over different zones. These servers act as cluster heads, which authenticates drones when entering a specific zone. This solution is, however, cost-inefficient as it requires installing multiple servers in each zone to be able to enforce the identification and authorization rules. Unlike the MCS solution, the distributed server solution is neither scalable nor cost-efficient. 
Finally, Allouch et al.} \cite{allouch2021utm} {proposed a permissioned blockchain to tackle the problem of mission planning and scheduling. Although they present an attractive protocol for managing UTM operations, the authors did not provide any mechanism that forces drone operators to act lawfully. 
In fact, none of the proposed systems provide a full policy enforcement mechanism that employs a reputation system and a dynamic pricing model to incentivize operators to follow the rules and penalize unlawful acts. }

\section{Security Analysis}\label{security}

\subsection{Vulnerability Analysis}
The Ethereum blockchain is a well-established platform that accommodates diverse applications. However, it is an antagonistic execution environment where attacks may exploit smart contracts' vulnerabilities to steal financial value. Therefore, a careful verification and checking of the smart contracts are crucial to avoid the monetary loss or rule-breaking. Particularly, malicious users may exploit any bugs in the UTM smart contract to initiate illegal missions, increase their reputation illegally, and cause air collisions or denial of airspace (DOAS). To make it easier for developer to inspect such bugs, a plenty of SC verification tools have been developed and published as open-source\cite{rameder2021systematic}. In our analysis, we consider four vulnerablity scanning tools, namely: SmartCheck \cite{tikhomirov2018smartcheck}, Oyente \cite{luu2016making}, Osiris \cite{ferreira2018osiris}, and Slither \cite{feist2019slither}. Table \ref{Verif} summaries the main features of these verification tools.


% \usepackage{multirow}
% \usepackage{multirow}
\begin{table*}[]
\centering
\caption{Verification tools used to detect security vulnerabilities in our smart contract }
\label{Verif}
\begin{tabular}{|c|c|c|c|}
\hline
\textbf{\begin{tabular}[c]{@{}c@{}}SC Verification \\ Tool\end{tabular}} & \textbf{Tool Description} & \textbf{\begin{tabular}[c]{@{}c@{}}Number of \\ scanned \\ vulnerabilities\end{tabular}} & \textbf{Outcome} \\ \hline
\textbf{SmartCheck \cite{tikhomirov2018smartcheck}} & \begin{tabular}[c]{@{}c@{}}Static analysis tool that \\ detects security, functional, operational,\\ and developmental issues.\end{tabular} & 20 & \multirow{4}{*}{\begin{tabular}[c]{@{}c@{}}No \\ vulnerabilities \\ detected in \\ our SC\end{tabular}} \\ \cline{1-3}
\textbf{Oyente \cite{luu2016making}} & \begin{tabular}[c]{@{}c@{}}Symbolic execution code that finds\\ security bugs in SC.\end{tabular} & 5 &  \\ \cline{1-3}
\textbf{Osiris \cite{ferreira2018osiris}} & \begin{tabular}[c]{@{}c@{}}A framework that employs \\ symbolic execution and taint analysis \\ to accurately identify integer bugs in SC.\end{tabular} & 3 &  \\ \cline{1-3}
\textbf{Slither \cite{feist2019slither}} & \begin{tabular}[c]{@{}c@{}}Static analysis tool that detects \\ vulnerable Solidity codes.\end{tabular} & 70 &  \\ \hline
\end{tabular}
\end{table*}

SmartCheck \cite{tikhomirov2018smartcheck} is a smart contract checking tool that performs static analysis to detect possible vulnerabilities in the solidity code. Basically, the tool is designed to detect a total of 20 vulnerabilities in solidity smart contracts which are categorized into security, functional, operational, and developmental vulnerabilities. This tool was tested on a large set of real-world contracts and was able to detect vulnerabilities in the majority of them. 
Our smart contract was successfully checked using the SmartCheck tool which identified no vulnerabilities. 

Oyente \cite{luu2016making} is, on the other hand, an alternative smart contract checking tool that is designed to analyze the low-level byte-code to detect issues such as transaction-ordering dependence, timestamp dependence, mishandled exceptions, and re-entrancy. We built a docker image of the tool and evaluated our smart contract inside a container. The report suggested that none of the five vulnerabilities are present in our code. 

Osiris \cite{ferreira2018osiris} is also a framework that employs symbolic analysis and taint analysis to detect integer bugs in smart contracts. It mainly focuses on detecting three types of integer bugs including, arithmetic bugs, truncation bugs, and signedness bugs. The tool identified none of these bugs in our smart contract. 

Finally, we tested our Solidity code using the Slither \cite{feist2019slither} verification tool. This tool is a static analysis tool that is designed to identify more than 70 security vulnerabilities with a false positive rate of 10.9\%. The tool checked our code successfully and reported no vulnerabilities.   

%\begin{figure}
 %   \centering
  %  \includegraphics[width = \linewidth]{Oyente.png}
   % \caption{Oyente security vulnerability report of our smart contract.}
    %\label{Oyente}
%\end{figure}

\subsection{Security Features}

To ensure security against cyber-attacks, it is vital to scrutinize the key security features of our solution. In this section, we carefully examine the confidentiality, integrity, availability, non-repudiation, and authentication of our proposed framework. We also highlight how our solution is designed to mitigate known attacks such as man-in-the-middle, denial-of-service, and replay attacks. 

\subsubsection{Confidentiality}
Confidentiality refers to protecting information from being accessed by unauthorized users. It is usually achieved by encrypting messages and data. By design, the Ethereum blockchain preserves confidentiality be means of public key infrastructure (PKI). Particularly, each user is assigned a unique identifier that is associated with pairs of asymmetric keys. As described in equation \ref{RI_D}, a verification code is issued by the USS for each mission which can only be verified by the USS and hence ensure the confidentiality of the UAV information as well as the mission plan. Especially in the context of crowdsensing, the flight plan information of the UAVs should not be revealed to the public, yet the public shall be able to receive and forward the RID to the authority smart contract. 

\subsubsection{Integrity}

Data integrity is important to ensure that transmitted data is not altered or modified by an intruder. The blockchain is a tamper-proof ledger that cannot be modified by any entity. The smart contract itself is as well tamper-proof, which means that no one can alter the pre-defined rules and regulations. The integrity of the blockchain is achieved by the employment of block hash. Changing any piece of data that is logged on the blockchain will require changing the hashes of all the other blocks in the chain. Together with the PKI, the block hash prevents the common man-in-the-middle attacks.  

\subsubsection{Availability}
This term refers to keeping the services provided by the UTM available to all users at all times. This requires special mechanisms to mitigate common attacks such as denial of service (DOS), denial of airspace (DOAS), and avoiding single-point-of-failure (SPF). The decentralized nature of the blockchain eliminates potential SPFs and DOS attacks because there is no central managing agent. Further, we introduced the dynamic mission fee to prevent possible DOAS attacks during peak hours. The dynamic fee increases as the number of drones flown increases and thus the cost of the DOAS attack becomes higher. We also limited the number of drone reports that can be sent by a certain user to one report per drone to prevent malicious reporters from affecting the reputation of a good drone. 

\subsubsection{Non-repudiation}

Non-repudiation is defined as the inability to deny or refuse responsibilities of actions. This can be achieved using PKI, where a UTM user signs messages using their private key before sending them via the network. Traceability and audibility features of the blockchain also ensure non-repudiation indirectly by logging all participants' activities on an immutable ledger that can be traced by all other participants. In addition, we ensure accountability of UAV operators by hashing the RID-VC which includes the mission plan and a nonce that is known only by the authority. In this way, we guarantee that the UAV operator is not able to reproduce the RID, while he is requested to broadcast it throughout the mission. This implicitly makes the UAV operator accountable for the mission plan that he received from the USS. He is also requested to broadcast the UAV's current location and timestamp as part of the RID. Finally, we employ crowdsensing as part of improving the non-repudiation feature of our solution. Particularly, MCS reports have deemed proof that a UAV was in a certain location at a certain time.

\subsubsection{Authentication and authorization}
Ensuring the authenticity of users and messages is a key requirement for the UTM. In principle, authentication is defined as the ability to recognize the real identity of a user. To achieve authentication, we employ access control modifiers for each function to make sure that it is called only by authorized users. 

\section{Limitations and Future Work}
Although our solution outperforms traditional WSN-based monitoring systems in terms of coverage and cost efficiency, it is worth noting that our design is still limited in some aspects that needs to be addressed in future work. Particularly, the MCS approach fails to identify drones that are not sending RID. For this particular case, a special detection and interdiction system needs to be put in place. Such system shall be able to identify the drone that is not sending its RID among many others in the sky, perform a risk assessment, and interdict high risk drones. Nonetheless, the deployment of such system will only complement the MCS-based system which will significantly reduce the overall deployment cost as compared to the traditional WSN-based approach. The integration of the two systems will be part of our future work. 

\section{Conclusion}\label{conclusion}

With the prevalence of UAV applications, the need for efficient and secure air traffic management solutions becomes inevitable. Despite national and international efforts towards regulated operation in the low-altitude airspace, decentralized solutions to enforce regulations and satisfy cybersecurity requirements are in high demand. This article described several issues with the current UTM system and proposed a novel solution to address these issues. This solution relies on a synergy between the concepts of blockchain smart contracts and mobile crowdsensing. The first is employed to regulate the access control to the airspace while ensuring high levels of confidentiality, integrity, availability, non-repudiation, and authentication. Mobile crowdsensing, on the other hand, is utilized as an efficient and scalable rule enforcement mechanism. An incentive mechanism was also presented to incentivize public users to report UAVs in the urban airspace. Our solution was implemented as two smart contracts on the Ethereum blockchain and verified using two security verification tools. A security analysis of the solution was provided to ensure compliance with general security requirements. Finally, a brief cost analysis was discussed to highlight the advantages of our solution as a business model. 

\section*{Acknowledgments}
This work is fully funded by the Center for Cyber-physical Systems at Khalifa University.



\begin{IEEEbiographynophoto}{Ruba Alkadi}
 received the B.Sc. degree in Electrical Engineering from the American Univesrity of Sharjah, Sharjah, UAE, in 2016, 
and the M.Sc. by research degree in engineering from Khalifa University, Abu Dhabi, UAE in 2018. She is currently a research associate in the Center of Cyber-Physical Systems in Khalifa Univesrsity. She is interested in the application of modern machine learning algorithms on big data analysis. 
\end{IEEEbiographynophoto}

\begin{IEEEbiographynophoto}{Abdulhadi Shoufan}
 received the Dr.-Ing. degree
from Technische Universität Darmstadt, Germany, in 2007. He is currently an Associate Professor of information security and electrical and computer
engineering and a member of the Center of Cyber Physical Systems, Khalifa University, Abu Dhabi. He is interested in drones’ security and safe operation as
well as in embedded security, learning analytics, and engineering education.
\end{IEEEbiographynophoto}

\vfill

\end{document}